\documentclass[11pt,twoside]{article}

\usepackage{asp2004,epsfig,aassym}

\begin{document}
\title{The GOODS-North Radio Galaxies: On the Origin of the Radio Emission}   %%% Fill in title
\author{G. Morrison, M. Dickinson, F. Owen, E. Daddi, R. Chary, F. Bauer, B. Mobasher, 
E. MacDonald, A. Koekemoer, A. Pope}   %%% Fill in author names
\affil{UHawaii-IfA/CFHT, NOAO, NRAO, NOAO, SSC, Columbia, STScI, NOAO, STScI, UBC}    %%% Fill in author affiliations

\begin{abstract} %%% Abstract to run on from here.
We report on a preliminary study concerning the origin of radio
emission within radio galaxies at L(1.4GHz)$>$1E24 W/Hz in the GOODS-N
field. In the local universe, Condon et al. (1989) and Yun et
al. (2001) have shown that in galaxies with radio luminosities greater
than $1\times 10^{23}$ W/Hz the majority of the radio emission
originates from a `monster' i.e., an AGN. Using the Chandra 2Msec
X-ray image centered on the GOODS-N field and a reprocessed VLA HDF
A-array data plus newly acquired VLA B-array data
($\sigma$=5.3$\mu$Jy), we find that radio galaxies (with spectroscopic
redshifts; all have z$>$1) with L(1.4GHz)$>$1E24 W/Hz typically have
an X-ray detection rate of 72\% (60\% emit hard X-rays suggesting an
AGN origin for the radio emission) in contrast to 25\% for radio
galaxies with $L < 1\times 10^{23}$ W/Hz. The ACS images of these
$L[1.4 GHz] \geq 1\times 10^{24}$ W/Hz galaxies typically show compact
rather than extended galaxy morphology which is generally found for
the less luminous radio emitting galaxies but a few appear to be
ongoing galaxy mergers. We also present SED fitting for these luminous
radio galaxies including Spitzer IRAC \& MIPS 24u photometry and 60\%
show distinct power-law SED indicative of an AGN. Initial results tell
us that the X-ray emitting radio galaxy population are generally not
submm sources but the few ($\sim$10\%) that are SCUBA sources appear
to be the small AGN population found by Pope et al.  and others.
\end{abstract}

%%% MAIN BODY OF TEXT GOES HERE. CONSULT "INSTRUCTIONS FOR AUTHORS USING
%%% LATEX2E MARKUP", SECTIONS 2.3-2.6 FOR HELP WITH EQUATIONS, FIGURES,
%%% AND TABLES.

\section{Introduction}   %%% Top level section head (remove "%" symbol)
Surveys of galaxies at $z > 1$ made from radio to X-ray (e.g., Owen \&
Morrison 2006, Hasinger et al. 1998, Lilly et al. 1996, Richards et
al. 1998) indicate that star-formation (SF) is an essential component
in the energetics of galaxies in the early universe.  Radio continuum
observations have been used as a sensitive, unbiased probe of
star-formation (SF) in galaxies to high redshift (e.g., Ivison et
al. 2002, Chapman et al. 2003). Using the well-known radio/far-IR
correlation (e.g., Dickey \& Salpeter 1984, Helou, Soifer
\& Rowan-Robinson 1985), the radio emission is linked to ongoing SF,
which is unaffected by dust obscuration. Radio observations also act
as a useful probe of AGN activity, despite any obscuration, through
radio morphology (requires sufficiently high resolution), spectral
index ($\alpha$, where $S_{\nu} \propto \nu^{\alpha}$), or via
anomalously high radio luminosities, $L \geq 1\times 10^{23}$
W/Hz. However the origin of the radio emission is not always easy to
determine from radio observations alone and at higher redshifts this
problem becomes more pronounced.  To detect the high-z AGN/starburst
(SB) population deep radio observations are required and to use the
spectral index to classify the radio galaxy population one must
observe at more than one frequency. Given the faintness of the $\mu$Jy
radio sources observing more that more frequency is prohibitive in
terms of integration time and because of the negative spectral index
at frequency higher than 1.4GHz. In addition, radio resolution (beam=
1.5\arcsec at 20cm) at high redshift is not sufficient to resolve the
radio morphology thereby yielding an ambiguity regarding the origin of
the radio emission. Thus while we know where to look for activity galaxies
but do not know which type of activity has been detected.

This preliminary study uses the excellent multi-wavelength data that
exist in the GOODS North field to assess the origin of the radio. In
the local universe, Condon et al. (2002) and Yun et al. (2001) have
shown that in galaxies with radio luminosities greater than $1\times
10^{23}$ W/Hz the majority of the radio emission originates from a
`monster' i.e., an AGN. Using the Chandra 2Msec X-ray image centered
on the GOODS-N field and a reprocessed VLA HDF A-array data plus newly
acquired VLA B-array data (rms=5.3$\mu$Jy), we determine which of the
L(1.4GHz)$>$1E24 W/Hz radio galaxies are hard X-ray sources the
signature of an AGN. 

%A variation of this method was used by Cowie et
%al. (2004) to separate out the AGN's from the radio sources so they
%could study the starburst population.

\begin{figure}[!ht]
\begin{minipage}[t]{1.0\linewidth}
\begin{minipage}[t]{.5\linewidth}
\includegraphics[width=5.cm, height=5.cm, angle=0]{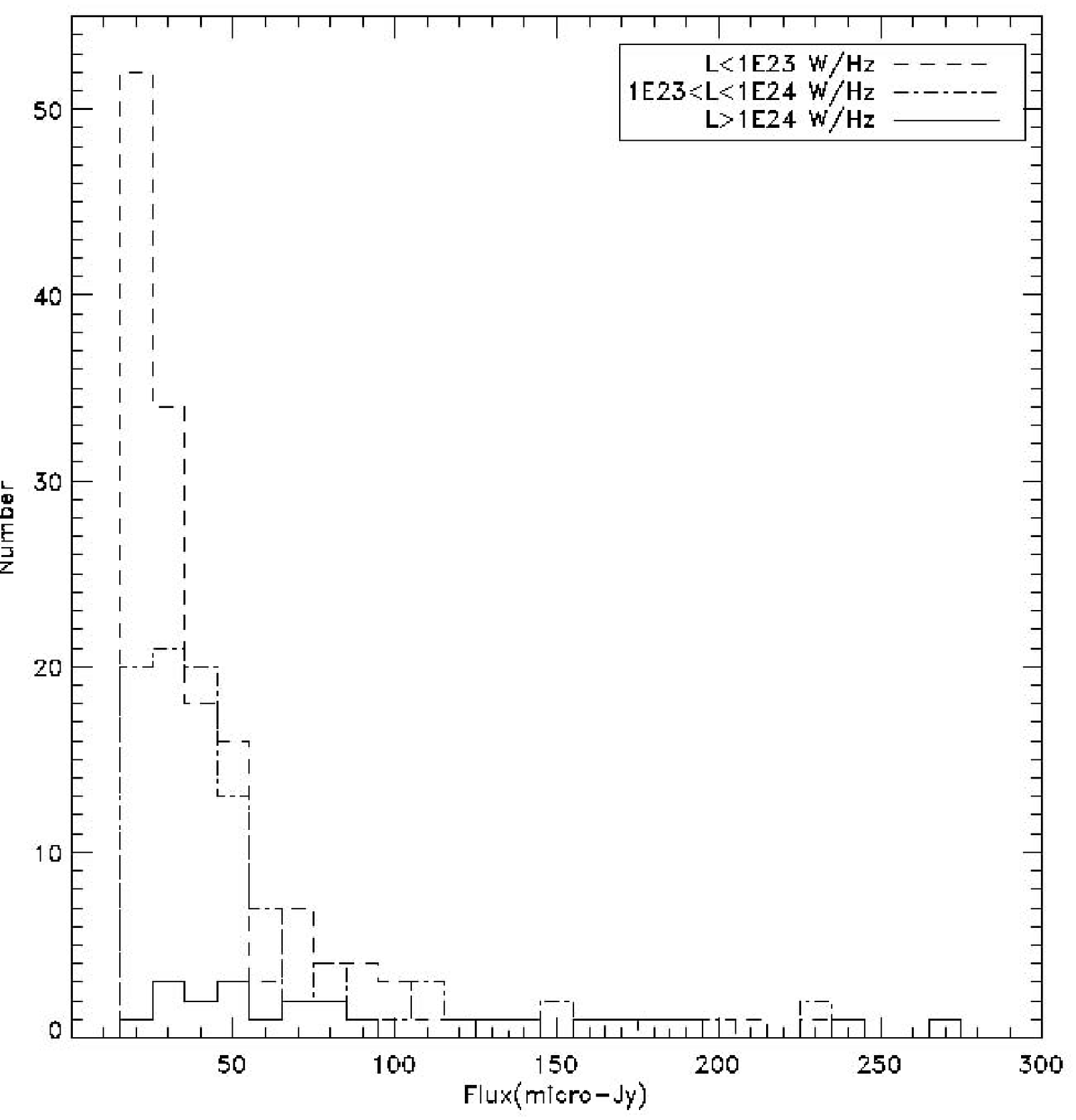}
\caption{Histogram of radio fluxes separated into
different radio luminosity classes using spectroscopic redshifts}\label{fig1}
\end{minipage}
\begin{minipage}[t]{.5\linewidth}
\includegraphics[width=5.25cm, height=5.25cm, angle=0]{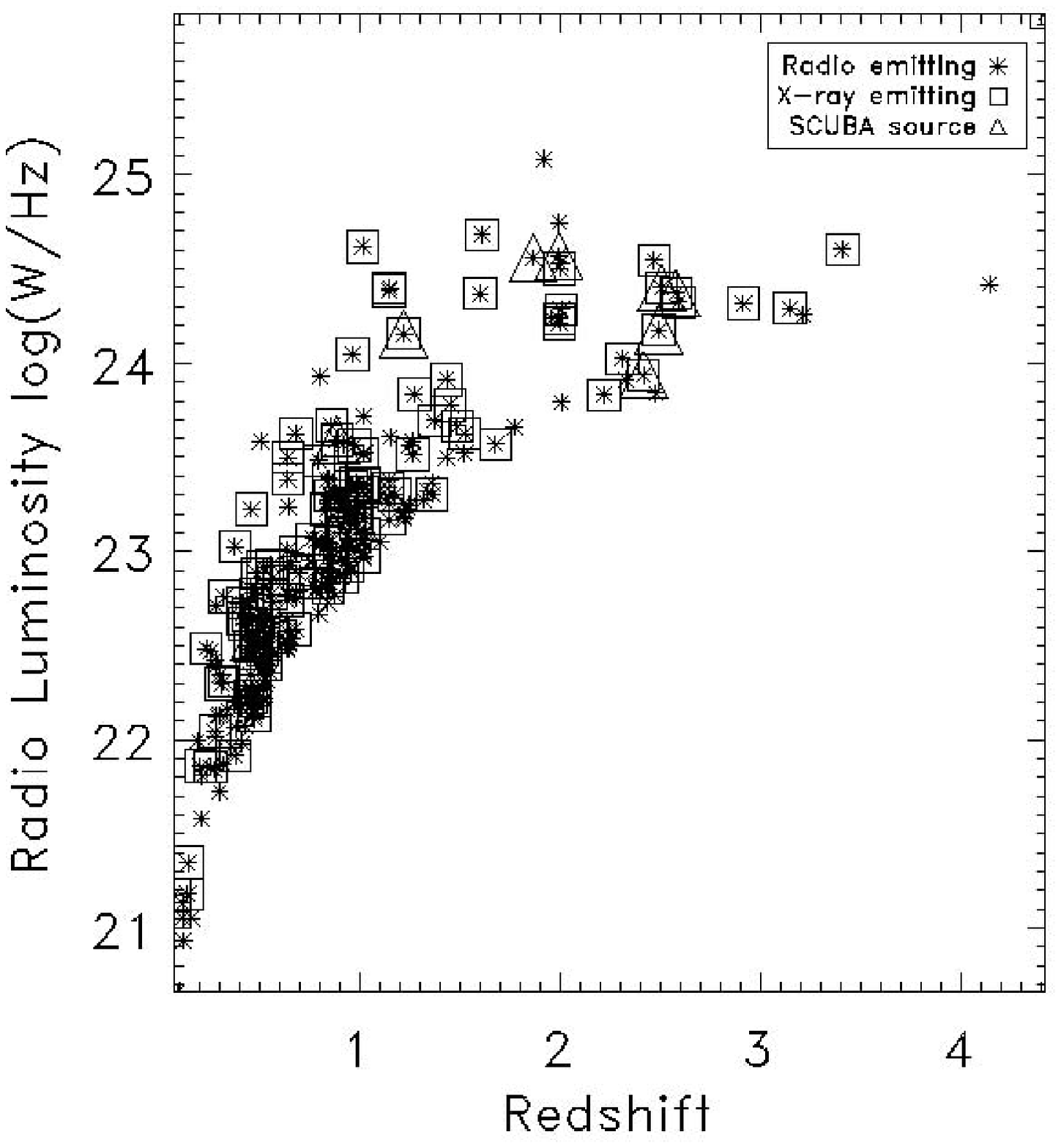}
\caption{Radio luminosity as a function of redshift. 2\,Msec Chandra
X-ray detections are denoted by squares and Scuba detection by
triangles.}\label{fig2}
\end{minipage}
\end{minipage}
\end{figure}

\section{Data}   %%% Lowest level section head (remove "%" symbol)
The X-ray data used in this analysis are from the ultra-deep Chandra
2Msec survey centered on the GOODS-N field. This survey resolves the
bulk of the 0.5-8.0 keV background, providing the deepest view of the
universe at this band (Alexander et al. 2003). The galaxian types
detected in this survey are absorbed and unabsorbed AGN-types and
starburst galaxies in decreasing order of source-density. Hard X-rays
(2-8 keV) are generally link to AGNs (Bauer et al. 2002) hence provide
a method for classifying the power source in radio galaxies.  The
radio data consists of 40hr of archival VLA A-array (Richards 2000)
and 28hr of new VLA B-array data centered on the GOODS-N field. AIPS
reprocessing of the archival data plus the addition of the new VLA
data has yielded a rms$\sim$5.3$\mu$Jy in the phase center of the
map. A 4$\sigma$ radio sources catalog was constructed using the AIPS
task SAD while the residual SAD image was inspected for missed
sources. Details can be found in Morrison et al. (2007). Further VLA
observations by Morrison et al. will total 150hr in A+B+C+D arrays
yielding a map with an estimated rms$\sim$3$\mu$Jy. These data will be
published in a future paper.  MIR \& FIR data were obtained from the
Spitzer Legacy project GOODS dataset which have obtained the deepest
observations with that telescope at 3.6-24$\mu$m using IRAC and MIPS
science detectors.  All radio galaxies in this study have measured
redshifts obtained from the literature. Information on the ground-base
optical/NIR \& HST-ACS data can be found in Giavalisco et
al. 2004. Submm data comes from the GOODS-N super-map (Borys et
al. 2003; Pope et al. 2005)

\begin{figure}[!ht]
\begin{minipage}[t]{1.0\linewidth}
\begin{minipage}[t]{.5\linewidth}
\includegraphics[width=5.cm, height=5.cm, angle=0]{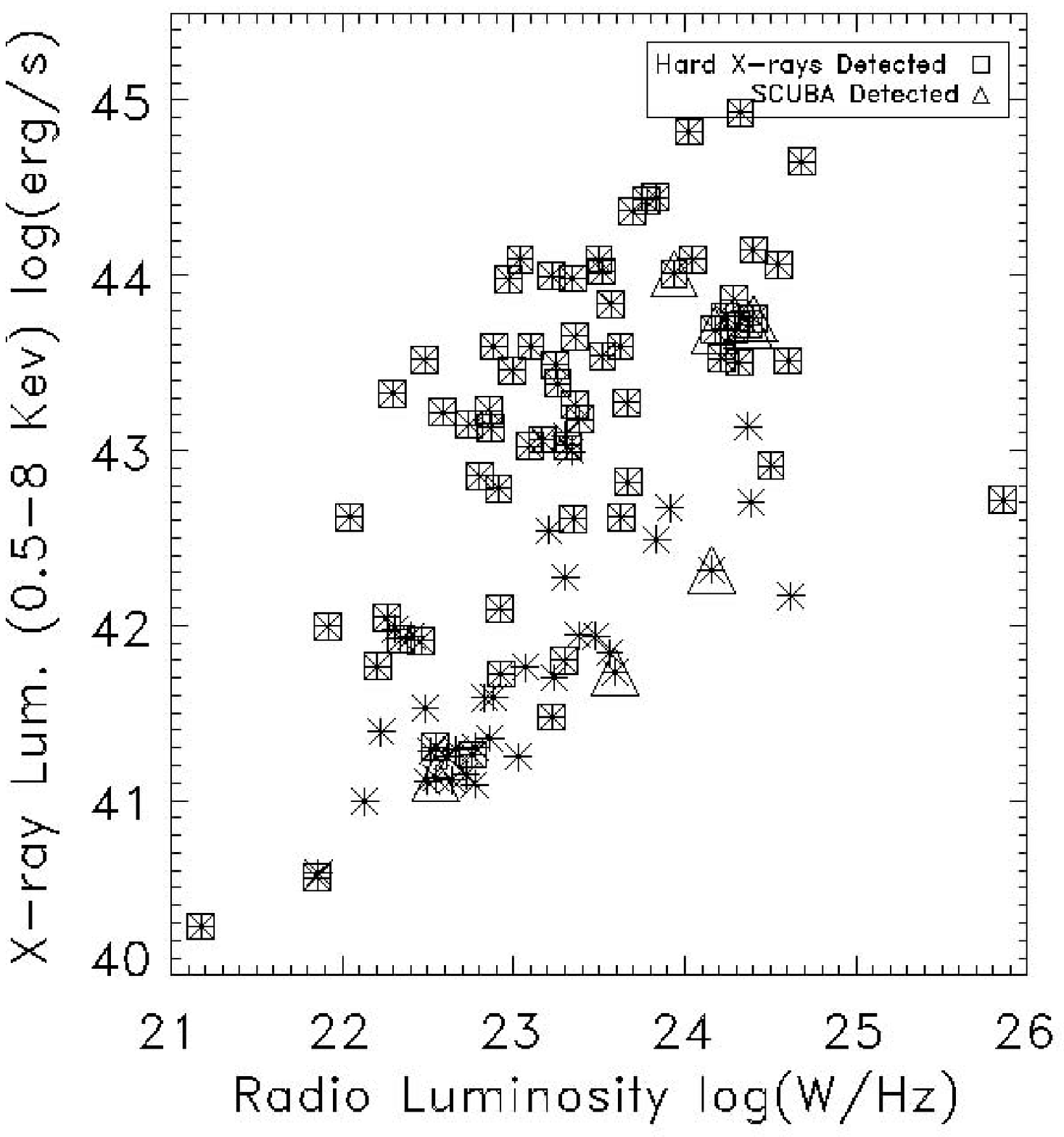}
\caption{Figure shows the radio luminosity and X-ray luminosity of the
72\% radio galaxies which are detected in the 2Msec Chandra data.}\label{fig3}
\end{minipage}
\begin{minipage}[t]{.5\linewidth}
\includegraphics[width=5.cm, height=5.cm, angle=0]{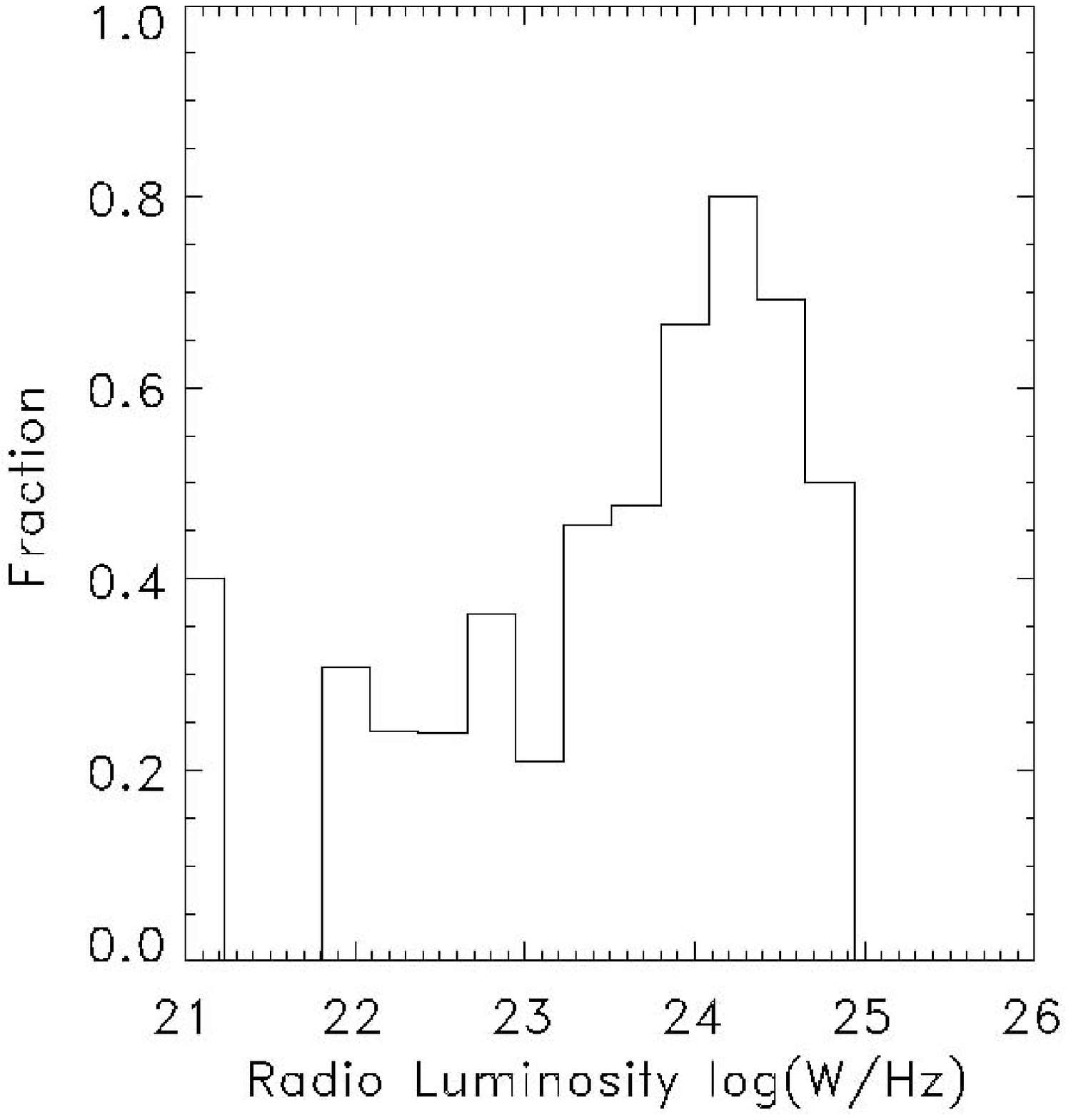}
\caption{Fraction 
of radio galaxies emitting X-rays (0.5-8KeV) normalized by the total
number of radio galaxies in a particular radio luminosity
bin.}\label{fig4}
\end{minipage}
\end{minipage}
\end{figure}

\section{Results}	%%% Unnumbered top level section head (remove "%" symbol)
Figure~\ref{fig1} is histogram of radio fluxes separated into
different radio luminosity classes using spectroscopic redshifts. At
least 10\% of the $\mu$Jy radio sources are made-up of radio galaxies
with $L \geq 1\times 10^{24}$ W/Hz. Here we consider the $L \geq
1\times 10^{24}$ W/Hz radio galaxy population.  In Figure~\ref{fig2}
we see the redshift distribution and the radio luminosity of the
galaxies. All radio galaxies in our sample with X-ray emission from
0.2-8.0 keV are label with squares (Alexander et al. 2003) while SCUBA
detected galaxies are denoted with a triangle (Pope et al. 2005). At
z$\sim$0, this radio population is made up of mainly AGN's (Yun et
al. 2001). The question we ask here is what powers the radio emission
at higher redshift: AGN or SB?  In Figure~\ref{fig2}, 72\% of the $L
\geq 1\times 10^{24}$ W/Hz radio galaxies are emitting X-ray and 80\%
of those are emitting hard (2-8KeV) X-rays (Fig.~\ref{fig3}) . Thus
~60\% of all the $L \geq 1\times 10^{24}$ W/Hz radio galaxies in this
sample are powered by an AGN if one assumes that hard X-ray are only
produced by AGNs. (However, it is known that many radio galaxies are
not X-ray luminous thus we are missing radio AGNs if we use only the
X-ray emission as a tracer of AGNs.)  Figure~\ref{fig4}, the fraction
of radio galaxies emitting X-rays (0.5-8KeV) normalized by the total
number of radio galaxies in a particular radio luminosity bin.  The
fraction varies from 25\% for $L < 1\times 10^{23}$ W/Hz and up to
72\% for $L \geq 1\times 10^{24}$ W/Hz. The turn over appears to be at
1$\times 10^{23}$ W/Hz the same radio luminosity Condon et al. (2002)
and Yun et al. (2001) found for the radio turnover from SB to AGN
powered radio emission.

We have done preliminary SED fitting for each galaxy in this sample. A
total of 13 photometric data points were used ranging from optical,
NIR, MIR,and 24$\mu$m.  Out of 29 radio galaxies in this sample with
$L \geq 1\times 10^{24}$ W/Hz, 17 or $\sim$60\% have a power-law
dominant SED.  Many of these galaxies have SED's that follow a
power-law from U-band through IRAC \& 24$\mu$m emission indicative a
AGN spectrum. The 2 SED fit uses the Bruzual/Charlot SEDs SEDs for the
optical/NIR portion of the SED while the MIR part is from Dale
\& Helou.  From Pope et al. (2005) $\sim$10\% of the radio emitting SCUBA
sources have hard X-ray emission thus these sources appear to be AGNs
based on X-ray diagnostics and their power-law SED.

The ACS images of these radio galaxies typically show compact rather
than extended galaxy morphology as is generally found for the less
luminous radio emitting SB galaxies. A few of these galaxies appear to
be ongoing galaxy mergers.
\section{Summary}
Our $L \geq 1\times 10^{24}$ W/Hz radio galaxy sample at $z>1$ has up
to an 80\% detection rate in X-rays and a 60\% detection rate in hard
X-rays indicating that for those radio galaxies their radio emission
is powered by an AGN. The fraction of detected radio galaxies in X-ray
varies from 25\% below 1$\times 10^{23}$ W/Hz to 80\% near $2\times
10^{24}$ W/Hz. The change in the X-ray detected fraction appears to
increase at 1$\times 10^{23}$ W/Hz the same radio luminosity region
where local radio galaxies transition between starburst powered radio
emission ($< 1\times 10^{23}$ W/Hz) and AGN powered radio emission
($\geq 1\times 10^{23}$ W/Hz). The radio galaxies with SCUBA
detections generally appear to be starburst since only a few have any
associated hard X-ray emission.  

\acknowledgements Support for GOODS, part of the Spitzer Space Telescope 
Legacy Science Program, was provided by NASA through Contract Number
1224666 issued by JPL, Caltech, under NASA contract 1407.

%%% Text of acknowledgments runs on after this command.

%%% THE BIBLIOGRAPHY
%%%
%%% CONSULT SECTION 3 OF "INSTRUCTIONS FOR AUTHORS" FOR HOW TO USE NATBIB.
%%% AUTHORS ARE ENCOURAGED TO USE EITHER THE "THEBIBLIOGRAPY" ENVIRONMENT
%%% BY UNCOMMENTING (DELETING THE "%" SYMBOL) THE COMMANDS BELOW, OR BY
%%% USING THE BIBTEX ENVIRONMENT. TO FIND OUT WHICH IS APPLICABLE TO YOUR
%%% CONTRIBUTION, CONSULT THE VOLUME EDITORS FOR YOUR PROCEEDINGS.
%%%

%\begin{thebibliography}{}
%\bibitem[]{}
%\bibitem[]{}
%\bibitem[]{}
%\bibitem[]{}
%\bibitem[]{}
%\bibitem[]{}
%\bibitem[]{}
%\bibitem[]{}
%\bibitem[]{}
%\bibitem[]{}
%\bibitem[]{}
%\bibitem[]{}
%\end{thebibliography}

\end{document}